\documentstyle[12pt,aasms4]{article}  
\def\lsim{\lower0.6ex\vbox{\hbox{$ \buildrel{\textstyle 
<}\over{\sim}\ $}}}
\def\gsim{\lower0.6ex\vbox{\hbox{$ \buildrel{\textstyle 
>}\over{\sim}\ $}}}

\def\beq{\begin{equation}}
\def\eeq{\end{equation}}
\def\beginapjbib{\begingroup \section*{\large \bf References}
         \parskip=.5ex plus 1.0pt
         \def\bibitem{\par \noindent \hangindent\parindent
                \hangafter=1}}
\def\endapjbib{\par \endgroup}
\def\alwaysmath#1{{\ifmmode{#1}\else{$#1$}\fi}}
\def\he#1{\hbox{\alwaysmath{{}^{#1}}{\rm He}}}

\def\hii{H\thinspace{$\scriptstyle{\rm II}$}~}
\def\Oii{O\thinspace{$\scriptstyle{\rm II}$}}
\def\Oiii{O\thinspace{$\scriptstyle{\rm III}$}}
\def\Nii{N\thinspace{$\scriptstyle{\rm II}$}}
\def\Sii{S\thinspace{$\scriptstyle{\rm II}$}}
\def\etal{{\it et al.}~}
\def\ie{{\it i.e.},~}
\def\eg{{\it e.g.},~}
\def\3he{$^3$He}
\def\4he{$^4$He}
\def\6li{$^6$Li}
\def\7li{$^7$Li}
\def\3h{$^3$H}
\def\Yp{Y$_{\rm P}$~}

\begin{document}

\vskip 0.7in
 
\begin{center} 

{\Large{\bf IONIZATION CORRECTIONS FOR LOW-METALLICITY \hii REGIONS AND 
THE PRIMORDIAL HELIUM ABUNDANCE}}
 
\vskip 0.4in
{Sueli M. Viegas$^1$, Ruth Gruenwald$^1$, and Gary Steigman$^2$}
 
\vskip 0.2in
{\it $^1${Instituto Astron$\hat{o}$mico e Geof\' \i sico, 
Universidade de S$\tilde{a}$o Paulo, \\
S$\tilde{a}$o Paulo, S.P. 04301-904, BRASIL}}\\

\vskip 0.1in 
{\it $^2${Departments of Physics and Astronomy,
The Ohio State University, \\ 
Columbus, OH 43210, USA}}\\
\newpage

{\bf Abstract}
\end{center}

Helium and hydrogen recombination lines observed in low-metallicity,
extragalactic \hii regions provide the data used to infer the
primordial helium mass fraction, Y$_{\rm P}$.  The ionization 
corrections for unseen neutral helium (or hydrogen) are usually 
assumed to be absent; \ie the ionization correction factor is 
taken to be unity ($icf \equiv 1$). In this paper we revisit the 
question of the $icf$ for \hii regions ionized by clusters of 
young, hot, metal-poor stars.  Our key result is that for the 
\hii regions used in the determination of Y$_{\rm P}$, there is 
a ``reverse'' ionization correction: $icf < 1$.  We explore the 
effect on the $icf$ of more realistic inhomogeneous \hii region 
models and find that for those regions ionized by young stars, 
with ``hard'' radiation spectra, the $icf$ is reduced further 
below unity.  In Monte Carlos using \hii region data from the 
literature (Izotov and Thuan 1998) we estimate a reduction in 
the published value of \Yp of order 0.003, which is roughly 
twice as large as the quoted statistical error in the \Yp 
determination.

\newpage

\noindent

\section{Introduction}

The primordial abundance of \he4 is key to testing the consistency of 
the standard hot big bang model of cosmology and to using primordial 
nucleosynthesis as a probe of cosmology and particle physics (for a 
recent review, see Olive, Steigman \& Walker 1999).  Since stars burn 
hydrogen to helium in the course of their evolution contaminating any 
primordial helium in the interstellar gas with their debris, it is the 
availability of large numbers of carefully observed, very low-metallicity 
extragalactic \hii regions which has permitted estimates of the primordial
helium mass fraction, $Y_{\rm P}$, whose statistical uncertainties are 
very small, $\approx$ 1\% (\eg see, Olive \& Steigman 1995 (OS), Olive, 
Skillman \& Steigman 1997 (OSS), Izotov, Thuan \& Lipovetsky 1994, 1997 
(ITL), Izotov \& Thuan 1998 (IT)).  However, along the path from the
observational data to the derived abundances, contamination by unknown 
systematic uncertainties may bias the inferred value of $Y_{\rm P}$.  
Observers have identified many potential sources of systematic error 
(Davidson \& Kinman 1985; Pagel \etal 1992 (PSTE); Skillman \etal 1994; 
Peimbert 1996; ITL; IT; Skillman, Terlevich \& Terlevich 1998) and, 
where possible, have designed their observing programs to minimize 
and/or to account for them.  In a previous paper (Steigman, Viegas \& 
Gruenwald 1997) we have explored the magnitude of the contribution to 
systematic error in \Yp from possible temperature fluctuations in the 
extragalactic \hii regions, concluding that they may have a significant 
effect on the determination of the primordial abundance of helium, 
comparable to or even greater than the statistical uncertainties.  
Here we turn our attention to another potential source of systematic 
error: the ionization correction for unseen neutral hydrogen and/or 
helium in \hii regions. 

Following Peimbert \& Costero (1969), an empirical method is usually 
employed to derive the abundances from the observed emission-line 
intensities.  The electron density and temperature of the gas are 
obtained from various emission-line intensity ratios and they are 
then used to calculate the appropriate line emissivities which, along 
with the observed emission line intensities, provide the fractional 
abundances of the various ions which are finally combined to obtain 
the element abundances (see, \eg Osterbrook 1989).  Since not all 
the ions present in the gas actually produce observable lines, an 
ionization correction must be applied to account for the ``missing" 
ionization states.  Historically, the ionization correction factor 
($icf$) has been derived for H, He, and many of the heavier elements 
from considerations of the ionization potentials or from numerical 
photoionization models (Peimbert and Torres-Peimbert 1977, Stasinska 
1980,1982, Mathis 1982, Pe\~na 1986).  In the quest for \Yp a key 
concern has been that due to its higher ionization potential, unseen 
neutral helium may lurk in those parts of the \hii regions where the 
hydrogen is still fully ionized ($icf$ $> 1$).  As the metallicity 
of the stellar population responsible for creating the \hii region 
decreases, the stars providing  the ionizing radiation are expected 
to be hotter, with ``harder" spectra.  If so, the helium and hydrogen 
ionization zones should be more nearly coincident and the ionization
correction minimized (\ie the $icf$ should be closer to unity).  

Some years ago, Pagel \etal (PSTE) proposed a method for estimating 
the helium $icf$ based on the ``radiation softness parameter" $\eta 
\equiv$ (0$^+$/S$^+$)(S$^{++}$/O$^{++}$) of Vilchez \& Pagel (1988).  
Comparing with photoionization models, they concluded that the $icf$ 
will differ negligibly from unity for log~$\eta < $ 0.9, corresponding 
to models with effective stellar temperatures higher than 37 000 K.  
When using a sample of \hii regions to derive Y$_{\rm P}$, PSTE impose 
this $\eta$ condition to discard from their data set those \hii regions 
for which the $icf$ may differ from unity (log~$\eta > $ 0.9) and 
they adopt $icf$ = 1 for all those \hii regions with log~$\eta < $ 
0.9.  In fact, as we shall see, for the sufficiently ``hard'' 
radiation field provided by a population of young, hot stars, 
the ionization correction may actually be reversed with neutral 
hydrogen present in those parts of the \hii regions where the 
helium is still ionized ($icf < $1) (Stasinska 1980,1982, Pe\~na 
1986, Dinerstein \& Shields 1986).  In either case, a potentially 
serious hidden assumption is that of homogeneity for the \hii 
regions (see, \eg Dinerstein \& Shields 1986).  Recent HST imaging 
reveals that these regions are far from homogeneous, showing many 
different features such as pillars, globules and even voids.  As 
a result the true $icf$ may differ significantly (on the scale of 
the statistical errors) from the values calculated in simplified 
models of homogeneous regions, introducing a systematic error in 
the calculation of the primordial abundance of \4he. For example, 
for \Yp $\approx 0.24$, if the $icf$ should differ from unity by 
1 -- 2\%, the change in \Yp will be of order 0.002 -- 0.004, 
comparable to the statistical uncertainties in \Yp suggested 
by the analyses of OS, OSS, and IT.  Furthermore, if indeed the 
$icf$ is less than unity for sufficiently young and metal-poor 
\hii regions, then even the seemingly conservative assumption 
of PSTE to adopt $icf$ = 1 may lead to a (metallicity-dependent) 
bias which systematically overestimates \Yp.

Another common simplification of the photoionization models used 
to estimate the $icf$ is the assumption that the ionizing radiation 
is provided by stars of a single stellar temperature.  In contrast, 
real \hii regions are ionized by clusters of stars of differing 
masses and temperatures whose ionizing radiation spectra differ 
from that of a single blackbody.  In the following we adopt the 
(time-dependent) spectrum of a starburst (Cid-Fernandes \etal 
1992) appropriate to low-metallicity stars and employ a numerical 
photoionization code (AANGABA; Gruenwald \& Viegas 1992) which 
allows us to account for inhomogeneity.  The models are described 
in \S 2.   The $icf$ results are discussed in \S 3 and are applied 
to an \hii region sample taken from the literature in \S4.  Our 
concluding remarks appear in \S5.

\section{Photoionization Models} 

Many independent 1-D photoionization codes have been developed 
over the last 30 years or so.  Comparison among several of them 
for some standard test cases of \hii regions, planetary nebulae, 
and active galactic nuclei reveals very good agreement (P\'equignot 
1986, Ferland \etal 1995).  These codes are in wide use analyzing 
observed emission-line spectra.  For such codes the usual input 
parameters are the ionizing radiation spectrum, the gas density, 
and the relative chemical abundances.  In most of the codes 
spherical or plane-parallel symmetry is assumed and the diffuse 
radiation is treated in the outward-only approximation.  Usually 
a constant density is assumed, or a specific functional form for 
the variation of density with distance is adopted.  However, 
these simplifying choices may not reflect the true structure 
of real \hii regions as revealed by HST imaging.  One way to 
approach the problem of more realistic modeling is to mimic 
the presence of condensations or voids by combining models 
for several different choices of input parameters.

In our analysis the AANGABA photoionization code is used to model 
the gas whose distribution is taken to be spherically symmetric.  
Later, in constructing more general inhomogeneous models, this 
restriction to spherical symmetry is relaxed. Because we are 
specifically interested in the helium/hydrogen $icf$ with the goal 
of analyzing its effect on the helium abundances derived from the 
low-metallicity \hii regions, a metal-poor chemical composition 
(0.1 solar) is chosen.  The stellar cluster ionizing radiation 
spectrum from Cid-Fernandes \etal (1992) is adopted and models are 
built for two evolutionary phases of the star cluster: the initial 
phase (t = 0), when the spectrum is dominated by the hottest, most 
massive stars, and a later phase (t = 2.5 Myr), when the hot stars 
have evolved and there are fewer He$^+$ ionizing photons.  In order 
to explore a variety of different physical conditions, the radiation 
intensity is characterized by the number of ionizing photons above 
the Lyman limit, Q$_H$.  Q$_H$ is directly related to the mass and 
IMF of the stellar cluster and, along with the gas density, defines 
the \hii regions.  In terms of the popular ionization parameter, U, 
for a given density each of our models corresponds to a fixed value 
of UR${_i}^{2}$ which is proportional to Q$_H$ (R$_i$ is the inner 
radius of the \hii region).  As a result, our constant density models 
should reproduce the $icf$ -- $\eta$ behaviour found in PSTE (their 
Figure 6).  Since the ionization parameter is proportional to Q$_H$/$n$, 
a grid of models at a given density and characterized by the value 
of the ionization parameter, corresponds to {\it different} values 
of Q$_H$.  It is true that those combinations of Q$_H$, n, and R$_i$ 
which keep U constant lead to the same ionization structure of the 
gas.  However, when looking for models to represent the observations 
of a specific \hii region, ionized by a given number of stars, any 
conclusions should be based on a grid with Q$_H$ constant.  Thus, 
in the following we show the results of a series of models with 
different values of U, satisfying the condition Q$_H$ = constant.  
First, for each choice of Q$_H$, a homogeneous (H) constant density 
model is constructed.

The above homogeneous models form the basis for the inhomogeneous models 
we create next: models with spatially limited density enhancements which 
will mimic density condensations (C) and those with density deficits for 
``valleys" (V).  These alternate models all start with the same input 
parameters as the corresponding homogeneous model.  We set the location 
of the condensations or valleys by specifying the corresponding optical 
depths at the Lyman limit, $\tau$, in the homogeneous model.  In our 
analysis we placed the condensations/valleys at the following locations: 
$\tau$ = 0.02, 0.03, 0.04, 0.10, 0.40.  At each of these radii the density 
(in a spherical shell) is either increased by a factor of 5 (condensation) 
or decreased by a factor of 2 (valley), extending over a distance which is 
10\% of the thickness of the \hii region for the corresponding homogeneous 
model.  To avoid numerical problems, the increase/decrease of the density 
followed a smooth analytical form.  Notice however that the size of this 
transition zone is always much smaller than the size of the condensation 
or the valley.  Aside from the condensation/valley, the density throughout 
the rest of the region is the same as in the corresponding homogeneous 
model.  We explored many other choices of these parameters but the ones 
adopted here provide a fair sample of the variations of the ionization 
correction factor (see the next section) in all the models we constructed.  
By constructing inhomogeneous \hii regions using as building blocks a 
variety of the H, C, and V models, we can mimic the physical conditions 
in realistic \hii regions using the Monte Carlo method.  

Examples of a different sort of composite model are those of Dinerstein 
\& Shields (1986), and of Pe\~na (1986) in which they account for the
contributions from superposed \hii regions with high and low stellar
temperatures (``hard'' and ``soft'' radiation fields).   In our analysis 
we can do the same, combining models for \hii regions using the stellar 
cluster ionizing radiation spectra at t = 0 and at t = 2.5~Myr.  In the 
following we pursue both kinds of inhomogeneous models.

\section{The Helium Ionization Correction Factor}

In \hii regions often only lines from the recombination of He$^+$ are 
observed, although occasionally those from the recombination of He$^{++}$ 
are also seen.  In order to account for the possible presence of any 
unseen He$^{++}$ and He$^0$, and also of H$^0$, we define the $icf$ 
so that the He/H abundance ratio is given by He/H $\equiv icf \times 
[n(He^{+})/n(H^{+})]$ so that,
\beq
icf = [1 + {(n(He^0) + n(He^{++})) \over n(He^+)}]/[1 + {n(H^0) \over n(H^+)}].
\eeq 
In addition, following Vilchez \& Pagel (1988), the radiation ``softness" 
parameter is defined as,
\beq
\eta = (n(0^+)/n(S^+))(n(S^{++})/n(O^{++})).
\eeq
Both the $icf$ and $\eta$ are obtained from the models described in the 
previous section.  

\subsection{Radiation Softness Parameter}

The use of the radiation softness parameter has been challenged by 
Skillman (1989). Using photoionization models characterized by the
value of the ionization parameter, he found that ``{\it for T$_{eff}$ 
$\geq$ 45,000 K, $\eta$ shows a strong dependence on U}".  This conclusion 
could invalidate the use of $\eta$ to obtain the helium ionization correction 
factor.  Skillman's models for {\it different} \hii regions are characterized 
by {\it different} values of the ionization parameter.  We note that Skillman 
uses the Dinnerstein \& Shields (1986) ionization parameter which is defined 
as the ratio beween the ionizing photon density and the gas density evaluated 
at the Str$\ddot{\rm o}$mgren radius when the inner radius is fixed at 10\% 
of the Str$\ddot{\rm o}$mgren radius (R$_{i}/R_{s} = 0.10$).  Each choice of
Skillman's ionization parameter corresponds to a {\it different} \hii region,
irradiated by a {\it different} number of ionizing photons Q$_H$, leading to
a {\it different} ionization structure.  It is, therefore, not surprising 
that when Skillman varies the ionization parameter he finds variations in 
the radiation softness parameter.  We have explored Skillman's claim using 
our homogeneous models and varying U (and R$_i$) while keeping Q$_H$ fixed.  
We recall that, as described in \S 2, each of our \hii region models is 
defined by {\it fixed} values of Q$_H$ for given choices of the gas density 
and filling factor and the ionization parameter (U) is defined as the ratio 
of the ionizing photon density to the gas density, evaluated at the inner 
radius.  We emphasize that the Q$_H$ range covered by our models is the 
same as that of Skillman's models.  In all cases we find that the models 
with {\it fixed} Q$_H$ but {\it different} U yield $\eta$ values which 
differ by less than 10\% and $icf$s which differ by less than 0.02\%.  
Exceptions do occur if the U value is too low, corresponding to a very 
large value of R$_i$.  These latter cases correspond to unrealistic models 
in that the geometrical width of the ionized region is much smaller than 
the inner radius (R$_s - $R$_i \ll~$R$_i$).  Nevertheless, even for these 
cases (low U, fixed Q$_H$) the $icf$ differs from that of the higher U 
models by less than 0.5\%.  Thus, if U is varied while keeping Q$_H$ 
(and the density) fixed, the radiation softness parameter $\eta$ is 
virtually unchanged, and there is a well-defined relation between the 
ionization correction factor and the Vilchez \& Pagel (1988) radiation 
softness parameter.

\subsection{Homogeneous Models}

First consider our results for the homogeneous models (the solid lines 
in Fig.~1).  For our fiducial models we fixed the density at n$_H$ = 
10 cm$^{-3}$ and varied Q$_H$ from 7.5 $\times$ 10$^{44}$ s$^{-1}$ to 
7.5 $\times$ 10$^{53}$ s$^{-1}$.  The lower bound on Q$_H$ is set by 
the requirement that the \hii region be bright enough to be observable.  
The arrows on the solid lines indicate the effect of increasing Q$_H$.  
By varying our choice of the Q$_H$ and the \hii region density we found 
that the solid (and dashed) lines are actually the loci of constant 
values of the combination $n^{2}$Q$_{H}$.  In addition, although our 
fiducial models assume a filling factor of unity ($\epsilon = 1$), we 
experimented with several choices of filling factor ($\epsilon < 1$) 
and verified that these models also lie along the lines in Figure~1, 
with a decrease in $\epsilon$ corresponding to an increase in 
$n^{2}$Q$_{H}$.  As noted in \S 3.1 above, the range in Q$_H$ 
covered by our models explores the same ionization parameter 
range that Skillman (1989) covered in his investigation.

As anticipated for the case where the nebula is ionized by a {\bf young} 
starburst with hot stars providing a hard radiation spectrum, the \hii 
region models predict a {\bf ``reverse"} ionization correction ($icf$ 
$\leq$ 1); \ie neutral hydrogen is present where the helium is still 
ionized.  As $n^{2}$Q$_{H}$ increases (and/or the filling factor 
decreases), $\eta$ decreases (from $\eta \la 5$ or log~$\eta \la 
0.7$) and the He$^+$ and H$^+$ regions more nearly coincide ($icf$ 
$\rightarrow$ 1) with the $icf$ -- $\eta$ relation being traced 
out by the solid line in the lower, left-hand corner of Figure~1.  

In contrast to the young starburst case, the softer spectrum of the 
ionizing radiation present 2.5~Myr after the initial burst leads to 
very different behavior (see the upper, right-hand corner of Fig.~1).  
In these cases neutral helium is present in the hydrogen ionized zone 
($icf$ $\geq$ 1).  At first, as Q$_H$ increases $\eta$ barely changes, 
decreasing only very slightly, while the $icf$  decreases noticeably
(in contrast to the young starburst case).  However, after turning the 
corner at the ``elbow'' located at log $\eta \approx 0.8$, $\eta$ then 
increases with increasing Q$_H$ while, now the $icf$ remains relatively 
constant.  

These distinct behaviors of \hii regions ionized by hard and by soft 
spectra were also found by Pe\~na (1986), by Dinerstein \& Shields 
(1986) and by PSTE.   Note in particular that for regions ionized by 
the soft spectra of ``old'' starbursts the radiation softness parameter, 
$\eta$, is bounded from below (log $\eta \ga 0.8$).  Since \hii regions 
with such large values of $\eta$ are normally excluded from analyses 
whose goal is the determination of the primordial abundance of helium, 
such old starbursts cannot dominate the ionizing radiation of the 
regions employed in such analyses.  Nonetheless, as noted by Dinerstein 
and Shields (1986), some contamination from regions ionized by such 
soft spectra could effect the overall ionization correction.  We 
explore this possibility below in \S 3.4.  

\subsection{Inhomogeneous Models}

Now consider the non-composite inhomogeneous models (C/V) whose 
construction was described in \S 2.  Since the results of the 
inhomogeneous ``valley" models are virtually indistinguishable 
from those of the corresponding homogeneous models, we have not 
shown them in Figure~1.  For the ``condensation" models we found 
that the greatest deviation from the corresponding homogeneous 
case occurs when the condensation is located at a radius equivalent 
to an optical depth of $\tau$ = 0.04.  At this distance the radiation 
field is still sufficiently strong to have been only partially 
absorbed by the high density gas and the regions shadowed by 
the condensation are still partially ionized, with the amounts 
of H$^0$ (for the high temperature cluster) and He$^0$ (for the 
low temperature cluster) determining the $icf$ value.  It is 
these cases ($\tau$ = 0.04) which are shown by the dashed lines 
in Figure~1.  Thus, all of our models, the non-composite C/V 
models as well as the composite models described below, lie 
between the homogeneous cases (solid lines) and these  
``extreme" inhomogeneous models (dashed lines).

For our Monte Carlos we used the H/V/C models as building blocks 
in the construction of a suite of composite inhomogeneous models. 
First, we fix the number of ionizing photons, Q$_H$, and the density, 
$n$, and archive the results of a set of corresponding H/V/C models 
with the valleys and condensations located at the different optical 
depths $\tau_{i}$ (see \S 2).  A single composite inhomogeneous 
model is created by assigning a weight proportional to the solid 
angle, $\Omega_i$, occupied by each H/V/C model and summing the 
weighted contributions of each model. The $\Omega_i$ are chosen 
randomly, subject to the constraint that their sum is 4$\pi$. 
This procedure is then repeated 10,000 times and the Monte 
Carlos are run for different choices of Q$_H$ and $n$.

Not surprisingly, the results of our Monte Carlos (for the young 
starburst case) all lie in the shaded region in Figure~1, between 
the $icf - \eta$ relation for the homogeneous models (solid curve) 
and that for the ``extreme" condensation ($\tau = 0.04$) models 
(dashed curve).  Although the behavior for the case of the old 
starburst is more complicated, it is clear that these composite 
inhomogeneous models will always have ``large'' values for the 
radiation softness parameter and cannot, by themselves, describe 
the \hii regions selected for probing the primordial helium abundance.

\subsection{Contamination By ``Old'' \hii Regions}

As we have seen, \hii regions ionized by young starbursts with hard
spectra always have a ``reverse'' ionization correction ($icf < 1$) 
and are limited to relatively low values of the radiation softness 
parameter (log $\eta \la 0.7$).  In contrast, regions ionized by a 
softer spectrum will have $icf > 1$, but also log $\eta \ga 0.8$.  
Although the latter would normally be eliminated from consideration 
by an $\eta$ cut, it is possible that a contribution from such 
regions could contaminate the helium abundances derived from 
observations (Dinerstein \& Shields 1986).  To test this possibility 
it would be necessary to construct models of superposed \hii regions 
for each observed \hii region and to constrain the parameters of the 
mixture by demanding that these composite models reproduce all the 
observed line strengths.  As an illustrative example, Dinerstein 
and Shields (1986) have done something simpler.  For NGC 4861 they 
superposed a ``hot'' region (T$_*$ = 55,000 K) and a ``cool'' region 
(T$_*$ = 35,000 K), fixing the ionization parameter for each region 
by the arbitrary requirement that each component reproduce the 
observed I($\lambda$6300)/I(H$\beta$) ratio.  The relative weights 
of each component were then fixed by the requirement that the resulting 
I($\lambda$3727)/I(H$\beta$) ratio have the observed value.  In this 
case they found that the low-T$_*$ component contributes 12\% of the 
observed H$\beta$ luminosity.  For this composite model they found 
$icf = 1.09$ (and log $\eta = 0.39$) in contrast to the value $icf = 
0.99$ (and log $\eta = 0.47$) for their simple, non-composite model, 
and they noted that ``{\it low observed values of} He$^{+}$/H$^{+}$
{\it cannot be interpreted as low} He/H {\it with complete confidence}''.  
However, it should be remarked that neither of their models is 
consistent with all their observed line ratios.  For example, neither 
their one-component nor their two-component model reproduces the 
observed value of the radiation softness parameter: log $\eta_{obs} 
= - 0.07$.  Notice from Figure 1 that for our one-component homogeneous 
and inhomogeneous models, the $icf \approx 0.99$ for log $\eta \approx 
- 0.07$.

As suggested by the failure of Dinerstein and Shields (1986) to 
account for all the observed line ratios, it may not be so easy 
to confuse a composite ``young/old'' \hii region with a single 
(albeit inhomogeneous) \hii region.  Various line ratios can be 
the key to distinguishing between the two possibilities.  Using 
our young (t = 0) and old (t = 2.5~Myr) \hii region models we have 
constructed a suite of composite models whose results are easy to 
understand with reference to Figure 1.  For almost all combinations
one or the other region will dominate and the $icf - \eta$ results 
will lie close to those of the dominant case (\ie the one with the 
largest Q$_H$).  Given the separation in log $\eta$ between the young 
and old regions, it is unlikely that an observed \hii region will 
be confused by a composite region dominated by a component with the 
``wrong'' value of log $\eta$.  Since virtually all of the \hii regions 
selected for determination of the primordial helium abundance have 
low values of log $\eta$ (see the data utilized in the next section), 
this suggests that {\it most} of the viable composite models will have 
$icf$s similar to those for our t = 0 case (\ie $icf \la 1$).  Only 
when the number of ionizing photons (Q$_H$) for each of the two regions 
(t = 0 and t = 2.5~Myr) are comparable may it be possible for a composite 
region to masquerade as a single region.  We have therefore concentrated 
on such ``democratic'' composite models and found that they occupy a 
region in the $icf$ -- log $\eta$ plane with log $\eta \ga 0.4$ and 
$icf \la 1.04$.  For the IT data which we use in our quantitative 
analysis in the next section, log $\eta \la 0.37$ and interloping 
regions are unlikely to have played a major role.  Nonetheless, we 
have pursued the question of whether such composite regions could 
contaminate the $icf$ inferred for \hii regions which are observed 
to have log $\eta \approx 0.4$.

As a first step we have compared the results of the t = 0, Q$_H$ = 
$7.5\times 10^{47}~s^{-1}$ model ($n = 10~$cm$^{-3}$ and $\epsilon 
= 1$) for which log $\eta = 0.37$ and $icf = 0.956$ with those for 
a composite model of two regions (t = 0 and t = 2.5~Myr) each with 
Q$_H$ = $7.5\times 10^{51}~s^{-1}$ ($n = 10~$cm$^{-3}$, $\epsilon 
= 1$) for which log $\eta = 0.42$ and $icf = 1.036$.  Notice that 
if $\epsilon < 1$, similar $\eta$ and $icf$ values are obtained 
with higher $n^{2}$Q$_{H}$.  Although the $\eta$s of the single 
and composite regions are very similar, we find that the individual 
ion ratios O$^{++}$/O$^{+}$ and S$^{++}$/S$^{+}$ differ by factors 
of 6 -- 7 between the simple and the composite models.  To explore 
this further, we compared the ratios of the intensities of various 
emission lines to that of H$\beta$ for the simple model (t = 0, 
Q$_H$ = $7.5\times 10^{47}~s^{-1}$) with those for a series of 
composite models with Q$_H$ varying from $7.5\times 10^{49}~s^{-1}$ 
to $7.5\times 10^{53}~s^{-1}$.  All of these composite models have 
log $\eta \approx 0.4$ and $icf \approx 1.04$.  In the composite 
models the [\Oii]($\lambda$3727) to H$\beta$ ratio varies from 18\% 
to 66\% of that in the simple model, the [\Nii]($\lambda$6584+6548) 
to H$\beta$ ratio is 16\% to 60\% of that in the simple model, and 
the [\Sii]($\lambda$6717+6731) to H$\beta$ ratio is 3\% to 41\% of 
that in the simple model.  In contrast, the [\Oiii]($\lambda$5007+4959) 
to H$\beta$ line ratios in the composite models exceed those in the 
simple model by factors of 1.4 to 2.3.  From our analysis it seems 
clear that these crucial line ratios can be key to unmasking composite 
\hii regions masquerading as simple \hii regions (see, also, Pe\~na 1986).

\section{Corrections To The Helium Abundance For Observed \hii Regions}

As we have done in our previous analysis of the effect of temperature
inhomogeneities on the primordial helium abundance inferred from the
observational data (Steigman, Viegas \& Gruenwald 1997), we quantify
the effect on \Yp of an $icf$ which differs from unity using a sample 
of \hii regions from the literature.  The data of Izotov and Thuan 
(IT) is especially well-suited to our task.  For 41 of the 45 \hii 
regions presented by IT, they provide helium abundances and data 
from which we may estimate the radiation softness parameter $\eta$.  
Although one of their \hii regions, UM 311, is not really in the 
category of ``metal-poor'', we have verified that whether or not 
we include it in our analysis has no significant effect on the 
{\it difference} in the derived value of \Yp using our calculated 
$icf$ as compared to the IT choice $icf = 1$ for all regions.  For 
the IT sample, $-0.25 \leq~ $log$~\eta~ \leq 0.37$ suggesting (see 
Fig.~1) an ``average'' $icf \approx 0.96 \pm 0.02$ and a potentially 
large reduction in Y$_{\rm P}$, $\Delta$\Yp $\approx - 0.0072 \pm 
0.0036$.  To quantify this reduction we have run a series of Monte 
Carlos using the IT data set and our computed $icf - \eta$ relations 
shown in Figure~1. 

For each \hii region in the IT sample we have $\eta$, the oxygen 
abundance O/H, and Y for $icf = 1$. For this fiducial data set 
we find the best linear fit to Y versus O/H in order to derive 
Y$_{\rm P}$.  We have also experimented with using averages of Y 
for the 15 -- 20 lowest oxygen abundance regions to bound \Yp and 
find the difference in our derived $\Delta$\Yp to be negligible 
compared to that found using the linear fit.  We emphasize that 
here we are not interested in the actual value of \Yp per se.  
Rather, we want to find the {\it reduction} in \Yp due to $icf 
\leq 1$.  In the Monte Carlos, for each of the observed regions 
we use the value of log $\eta$ from the data to randomly choose 
an $icf$ between the minimum value for a condensation located at 
$\tau = 0.04$ (``extreme'' inhomogeneous case) and the $icf$ for 
the homogeneous case.  In choosing between these two extremes we 
have experimented with three probability distributions: equal 
probability for the $icf$ to lie between the two extremes and 
decreasing/increasing probability for the extreme inhomogeneity.  
For each of the (41) IT regions we correct their value of 
He$^{+}$/H$^{+}$ with the randomly chosen $icf$ and compute the 
$icf$-corrected He$^{+}$/H$^{+}$ ratio which we use to find the 
$icf$-corrected Y. This new set of (41) Y, O/H pairs is used to 
find the corresponding \Yp from the linear Y versus O/H fit.  For 
each choice of $icf$ probability distribution we redo this 10,000 
times.  Our results appear as histograms for the distributions 
of $\Delta$\Yp in Figure~2.  It is clear from Figure 2 that our 
principal conclusion, $-\Delta$\Yp $\approx  0.003 - 0.004$, does 
not depend on the choice of $icf$ probability distribution.  Our 
Monte Carlos thus suggest that by ignoring the correction for 
He -- H ionization, IT have {\it over}estimated the primordial 
helium mass fraction by 0.0034 $\pm$ 0.0003.  For their sample, 
IT derive (see their Table 7) \Yp(IT) = 0.2443 $\pm$ 0.0015 (or, 
0.244 $\pm$ 0.002).  Including our estimate of the true $icf$ we 
suggest their value of \Yp should be reduced to Y$_{\rm P}$(VGS) 
= 0.241 $\pm$ 0.002.  We note that $icf \equiv 1$ has also been 
assumed for the \hii regions in the data sets used by OS and OSS, 
so their estimates of \Yp should likely be reduced by a similar 
amount to that which we have found for the IT sample.

\section{Conclusions}

Relatively accurate helium abundance determinations are currently 
available for nearly 100 low-metallicity extragalactic \hii regions
(see, \eg OS; OSS; IT).  These data enable estimates of the primordial 
abundance of helium to unprecedented statistical accuracy.  For 
example, IT quote $\Delta$\Yp = $\pm$ 0.0015 (or, $\pm$ 0.002) 
for the regions they have observed and analyzed.  At this level 
of uncertainty unanticipated systematic errors may overwhelm the 
statistical uncertainties (see, \eg Davidson \& Kinman 1985; PSTE; 
Skillman \etal 1994; ITL; Peimbert 1996; Steigman, Viegas \& Gruenwald 
1997; Skillman, Terlevich \& Terlevich 1998; IT). In this paper we 
have revisited the question of the helium ionization correction and 
its contribution to estimates of Y$_{\rm P}$.  Virtually all the 
low-metallicity \hii regions employed in the quest for \Yp have 
``hard'' spectra (low $\eta$) and, almost always the ionization 
correction has been ignored or, rather, {\it assumed} to be unity 
(no correction).  We have seen, as have others before us (\eg 
Pe\~na 1986, Dinerstein \& Shields 1986; PSTE), that there is a 
{\it reverse} ionization correction ($icf < 1$) for regions ionized 
by such hard spectra.  Inhomogeneities, surely present in these 
regions, only serve to exacerbate this correction (\ie to reduce 
the $icf$).  In our comparison with real data (IT), we suggest 
that their estimate of \Yp = 0.244 $\pm$ 0.002, should be reduced 
to \Yp = 0.241 $\pm$ 0.002, a correction {\bf larger} than the 
quoted statistical error.  Of course, this latter estimate still 
ignores other possible sources of systematic error such as temperature 
fluctuations (Peimbert 1971; Steigman, Viegas \& Gruenwald 1997) 
and underlying stellar absorption (see ITL; Skillman, Terlevich 
\& Terlevich 1998; IT).

In the current precision era of cosmology predictions and observations
are achieving unprecedented levels of statistical accuracy.  This is
certainly the case for primordial nucleosynthesis in the standard
(isotropic, homogeneous, three flavors of light neutrinos, etc.) hot
big bang cosmological model (BBN).  For example, at a fixed value of
the baryon density (or, equivalently, the baryon-to-photon ratio) the
uncertainty in the predicted primordial abundance of helium is at the
level of $\pm$ 0.0005 (Hata \etal 1996; Burles \etal 1999; Olive,
Steigman \& Walker 1999; Lopez \& Turner 1999).  Furthermore, the
variation of Y$_{\rm P}$(BBN) with baryon density is very slow so that 
if deuterium is used as a baryometer to constrain the baryon density, 
$\Delta$Y$_{\rm P}$(BBN) $\approx 0.006(\Delta y_{2}/y_{2}$) where 
$y_{2} \equiv$ (D/H)$_{\rm P}$.  The uncertainty in the BBN-predicted 
primordial deuterium abundance is of order 8\% (Hata \etal 1996) 
or less (Burles \etal 1999), a level comparable to that claimed 
for the uncertainty in the value inferred from observations of 
deuterium along the lines-of-sight to two high-redshift, low-metallicity 
(hence, very nearly primordial) QSO absorption line systems (Burles 
\& Tytler 1998a,b): $y_{2}(obs) = 3.4 \pm 0.25 \times 10^{-5}$.  The 
BBN helium abundance which corresponds to this deuterium abundance 
is Y$_{\rm P}$(BBN) = 0.247 $\pm$ 0.001.  Notice that while this 
expected abundance is consistent with the IT determined value (0.244 
$\pm$ 0.002) it is higher than our IT-corrected estimate (0.241 
$\pm$ 0.002) by some 3$\sigma$.

Strictly speaking, our quantitative estimate of the reduction in 
\Yp associated with the ionization correction factor applies only
to the data of IT.  Nonetheless, it is anticipated that \Yp inferred 
from the data assembled by OS should be similarly reduced.  Since 
the OS estimate of \Yp (excluding the NW region of IZw18 which is 
suspected of being contaminated by underlying stellar absorption) 
is even lower than that of IT (Y$_{\rm P}$(OS) = 0.234 $\pm$ 0.003) 
the discrepancy between theory and data is even larger.  Of course 
it could be that the Burles and Tytler (1998a,b) estimate of the 
primordial deuterium abundance is too low (see, \eg Webb \etal 
1997; Levshakov, Kegel and Takahara 1998a,b, 1999).  Nonetheless, 
this example provides an object lesson highlighting the importance 
of careful estimates of systematic errors in the current era of 
precision cosmology.

In deriving the helium abundance from observations of \hii regions it 
is necessary to correct the emission-line data for various effects (\eg 
collisional excitation).  In this paper we have explored the correction 
for unseen hydrogen and/or helium using detailed photoionization models 
of \hii regions ionized by realistic radiation spectra from different 
(young/hot and old/cool) stellar clusters.  For regions ionized by young, 
hot stars the radiation softness parameter is small (log~$\eta~ \la 0.7$) 
and there is a systematic, {\it reverse} ionization correction ($icf < 1$).  
In contrast, for regions ionized by an older stellar population with fewer 
hot stars the radiation softness parameter is large (log~$\eta~ \ga 0.8$) 
and $icf > 1$.  Thus, the correction for ionization is systematically 
correlated with the magnitude of the radiation softness parameter.  Of 
course, if this correction were very small it would have a negligible 
effect on the helium abundance determination and could be neglected.  
Our results suggest this is not the case.  For the \hii regions in the 
IT data set there are {\it reductions} in individual Y values which could 
be as large as 0.01 or even larger.  These {\it reductions}, which are 
comparable to (or, even larger than) typical statistical errors in Y 
for individual \hii regions (see, \eg OS, OSS, IT), should be compared 
to the theoretical (SBBN) uncertainty which is more than one order of 
magnitude smaller(\eg Olive, Steigman and Walker 1999).  For the IT 
data set we find the reduction in the inferred primordial mass fraction 
is comparable to (or, even somewhat larger than) the quoted statistical 
error in Y$_{\rm P}$.  The importance of including the ionization 
correction may be illustrated by the corresponding bound on N$_{\nu}$, 
the ``effective" number of light neutrino species, which provides a 
measure of any deviation from the standard model energy density at the 
epoch of BBN (Steigman, Schramm and Gunn 1977).  For SBBN, N$_{\nu} = 3$, 
while the presence of ``new", light particles (beyond the standard model 
of particle physics) would permit N$_{\nu} > 3$.  In a recent analysis 
utilizing the IT determination of \Yp and the Burles and Tytler (1998a,b) 
deuterium, Burles \etal (1999) find a 2$\sigma$ upper bound of N$_{\nu} 
\leq 3.2$ (see, also, Olive, Steigman and Walker 1999).  Incorporating 
the {\it reduction} in \Yp from the ionization correction will reduce 
this bound to N$_{\nu} \leq 2.9$, posing a potential challenge to SBBN.  
While there undoubtedly remain other, as yet unquantified, systematic 
errors whose magnitude may be larger than the ionization correction, 
it seems clear that the systematic reduction for ionization cannot be 
ignored.

\vskip 0.5truecm

\noindent {\bf Acknowledgments}

\vskip 0.5truecm

In Brazil the work of S.M.V. and R.G. is partially supported by grants 
from CNPq (304077/77-1 and 306122/88-0), from FAPESP (98/02816-6), 
and from PRONEX/FINEP (41.96.0908.00); in the U.S. the work of G.S. 
is supported at The Ohio State University by DOE grant DE-AC02-76ER-01545.  
Some of this work was done while S.M.V. was visiting the OSU Physics 
Department and while G.S. was visiting IAGUSP and they wish to thank 
the respective host institutions for hospitality.  

\newpage

\vskip 0.5truecm

\beginapjbib

\bibitem Burles, S. \& Tytler, D. 1998a;b, ApJ, 499, 699; {\it ibid}
507, 732

\bibitem Burles, S., Nollett, K.M., Truran, J.M., \& Turner, M.S. 1999,
preprint (astro-ph/9901157)

\bibitem Cid-Fernandes, R., Dottori, H., Gruenwald, R., \& Viegas, S.M. 
1992, MNRAS, 255, 165

\bibitem Davidson, K. \& Kinman, T.D. 1985, ApJS, 58, 321

\bibitem Dinerstein, H.L. \& Shields, G.A. 1986, ApJ, 311, 45

\bibitem Ferland, G. \etal 1995, in The Analysis of Emission Lines,
STScI Symp. Ser. 8, ed. R. E. Williams \& M. Livio (Cambridge;
Cambridge University Press), 143

\bibitem Gruenwald, R. \& Viegas, S.M. 1992, ApJS, 78, 153

\bibitem Hata, N., Scherrer, R.J., Steigman, G., Thomas, D., \& Walker, 
T.P.  1996,  ApJ, 458, 637

\bibitem Izotov, Y.I., Thuan, T.X., \& Lipovetsky, V.A. 1994
ApJ 435, 647 (ITL)
 
\bibitem Izotov, Y.I., Thuan, T.X., \& Lipovetsky, V.A. 1997, 
ApJS, 108, 1  (ITL)

\bibitem Izotov, Y.I. \& Thuan, T.X. 1998, ApJ 500, 188 (IT)

\bibitem Levshakov, S.A., Kegel, W.H., \& Takahara, F. 1998a, ApJ, 499, L1 

\bibitem Levshakov, S.A., Kegel, W.H., \& Takahara, F. 1998b, A\&A, 336, L29

\bibitem Levshakov, S.A., Kegel, W.H., \& Takahara, F. 1999, MNRAS, 302, 707

\bibitem Lopez, R. \& Turner, M.S. 1999, Phys. Rev. D56, 103502

\bibitem Mathis, J.S. 1982, ApJ, 261, 195

\bibitem Olive, K.A., Skillman, E. \& Steigman, G. 1997, ApJ, 483, 788 (OSS)

\bibitem Olive, K.A., \& Steigman, G. 1995, ApJS, 97, 49 (OS)

\bibitem Olive, K.A., Steigman, G., \& Walker, T.P. 1999, Physics Reports, 
in press (astro-ph/9905320)

\bibitem Osterbrock, D. 1989, Astrophysics of Gaseous Nebulae and 
Active Galactic Nuclei (Mill Valley: University Science Books)

\bibitem Pagel, B.E.J., Simonson, E.A., Terlevich, R.J.
\& Edmunds, M. 1992, MNRAS, 255, 325 (PSTE)

\bibitem Peimbert, M. 1971, Bol. Obs. Tonantzintla y Tacubaya, 6, 29

\bibitem Peimbert, M. 1996, Rev. Mex. Astr. Astrofis., Serie de 
Conferencias, 4, 55 

\bibitem Peimbert, M. \& Costero, R. 1969, Bol. Obs. Tonantzintla 
y Tacubaya, 5, 3

\bibitem Peimbert, M \& Torres-Peimbert, S. 1977, MNRAS 179, 217

\bibitem Pe\~na, M. 1986, PASP, 98, 1061

\bibitem P\'equignot, D.  1986, Workshop on Model Nebulae, ed. D. 
P\'equignot (Meudon: Observatorie de Paris-Meudon)

\bibitem Skillman, E.D. 1989, ApJ 347, 883 

\bibitem Skillman, E., Terlevich, R.J., Kennicutt, R.C., Garnett, D.R., 
\& Terlevich, E. 1994, ApJ, 431, 172

\bibitem Skillman, E.D., Terlevich, E., \& Terlevich, R.J. 1998, Sp. Sci. 
Rev., 84, 105

\bibitem Stasinska, G. 1980, A\&A, 84, 320

\bibitem Stasinska, G. 1982, A\&A Suppl., 41, 513

\bibitem Steigman, G., Schramm, D.N., \& Gunn, J.E. 1977, Phys. Lett. B, 
66, 202 

\bibitem Steigman, G., Viegas, S.M., \& Gruenwald, R. 1997, ApJ, 490, 187

\bibitem Vilchez, J.M. \& Pagel, B.E.J. 1988, MNRAS, 231, 257

\bibitem Webb, J.K., Carswell, R.F., Lanzetta, K.M., Ferlet, R., 
Lemoine, M., Vidal-Madjar, A., \& Bowen, D.V. 1997, Nature, 388, 250

\endapjbib

\newpage

\noindent{\bf{Figure Captions}}

\vskip.3truein

\begin{itemize}

\item{{\bf  Figure 1:} The $icf - $log$~\eta$ relations for \hii regions
ionized by ``young'' (t = 0) and ``old'' (t = 2.5~Myr) star clusters.  
The results for the young star clusters occupy the lower, left-hand 
part of the figure, while those for the older star clusters are found 
in the upper, right-hand part of the figure.  The solid lines are for 
the homogeneous models, while the dashed lines are for the ``extreme'' 
inhomogeneous models (see the text).  The shaded region between the 
solid and dashed lines for the t = 0 case are the results of Monte 
Carlos for composite inhomogeneous models.} 

\item{{\bf Figure 2:} Distribution of the {\it offsets} in the primordial 
helium mass fraction, Y$_{\rm P}$, inferred from Monte Carlos using 
the IT data and our $icf - \eta$ results for three different probability 
distributions which interpolate between the homogeneous and ``extreme'' 
inhomogeneous models (see the text).  From top to bottom the probability 
distributions favor the homogeneous models, are neutral, and favor the 
inhomogeneous models.} 

\end{itemize}

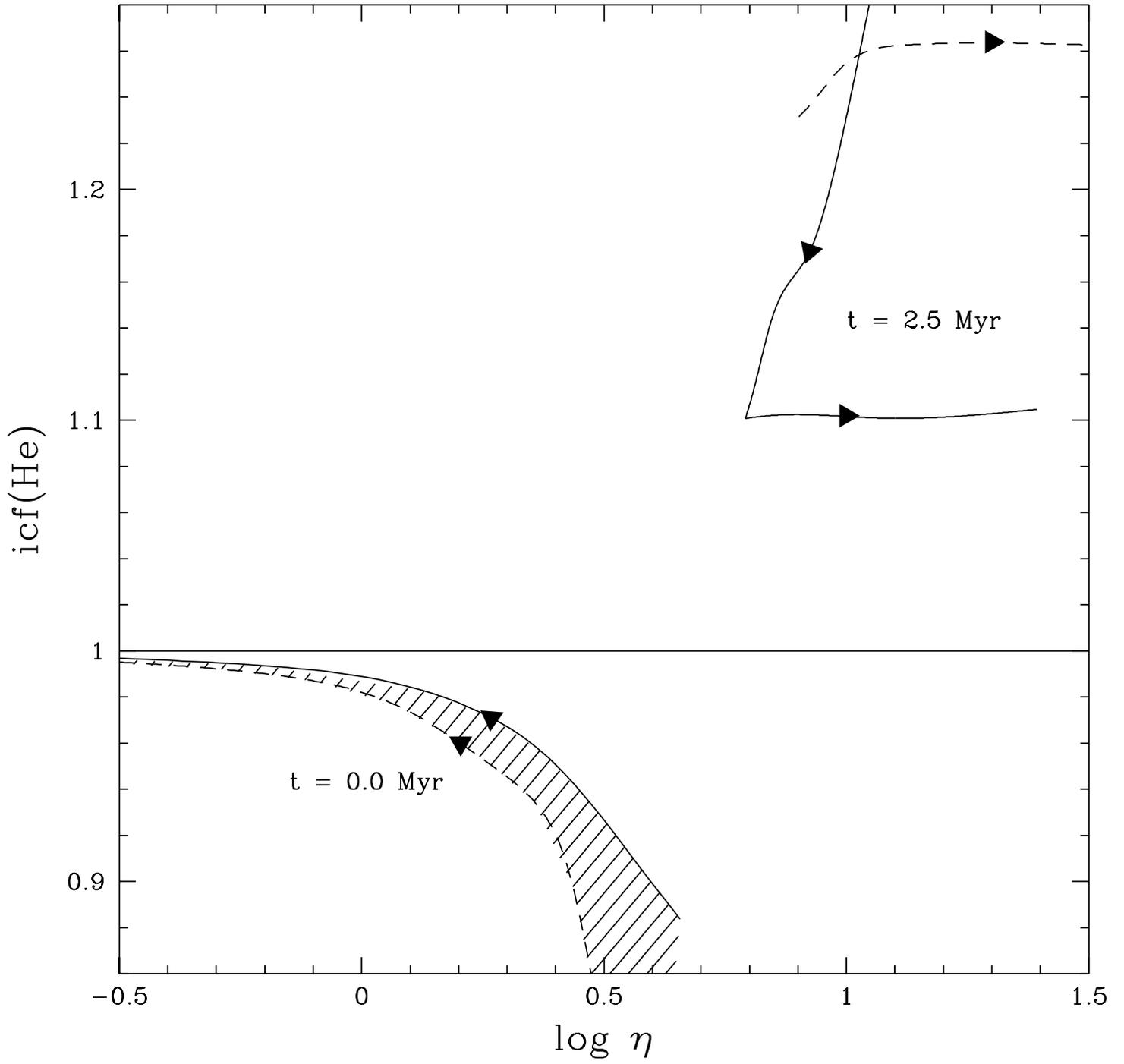
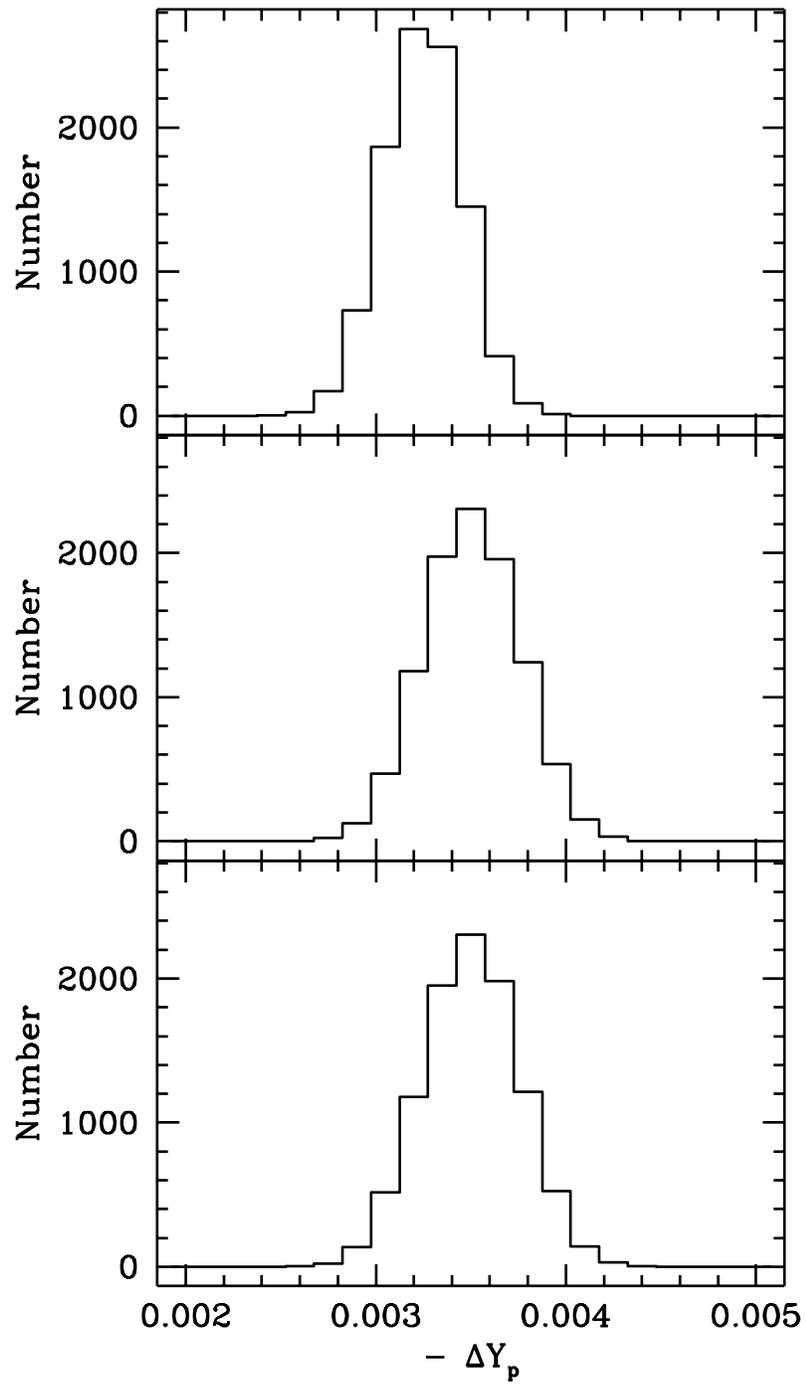
\end{document}